\documentclass[a4paper]{article}

\usepackage{spconf}
\usepackage{amssymb}
\usepackage{url}
\usepackage{amsmath}
\usepackage{booktabs}
\usepackage{graphicx}
\usepackage{threeparttable}
\usepackage[ruled,vlined]{algorithm2e}
\usepackage{float}

\usepackage[usenames,dvipsnames]{xcolor}
\usepackage{xcolor}
\SetNlSty{textbf}{\color{black}}{}

\SetCommentSty{mycommentfont}

\def\email#1{\gdef\@email{#1}}
\gdef\@email{email@address}

\title{A Study of Transducer based End-to-End ASR with ESPnet:\\ Architecture, Auxiliary Loss and Decoding Strategies}
\name{Florian Boyer$^{1,2}$, Yusuke Shinohara$^3$, Takaaki Ishii$^3$, Hirofumi Inaguma$^4$, Shinji Watanabe$^{5,6}$}
\address{
  $^1$ Airudit, Speech Lab., $^2$ LaBRI, Bordeaux INP, CNRS, UMR 5800, $^3$ Yahoo Japan Corporation,\\$^4$ Kyoto University, $^5$ Carnegie Mellon University, $^6$ Johns Hopkins University\vspace{-0.1cm}}

\begin{document}
\ninept
\maketitle
\begin{abstract}

In this study, we present recent developments of models trained with the RNN-T loss in ESPnet. It involves the use of various architectures such as recently proposed Conformer, multi-task learning with different auxiliary criteria and multiple decoding strategies, including our own proposition. Through experiments and benchmarks, we show that our proposed systems can be competitive against other state-of-art systems on well-known datasets such as LibriSpeech and AISHELL-1. Additionally, we demonstrate that these models are promising against other already implemented systems in ESPnet in regards to both performance and decoding speed, enabling the possibility to have powerful systems for a streaming task. With these additions, we hope to expand the usefulness of the ESPnet toolkit for the research community and also give tools for the ASR industry to deploy our systems in realistic and production environments.
\end{abstract}

\section{Introduction}

In recent years, end-to-end models based on either CTC \cite{Graves06-CTC}, attention encoder-decoder \cite{Bahdanau15-ETE}, or RNN-Transducer \cite{Graves12-STW} have gained a lot of attention from the speech recognition community, surpassing traditional hybrid ASR systems on various speech recognition tasks. Among them, transducer\footnote{We refer to "transducer" from here due to the different architecture.} was found successful in both research and industry environments for its competitive results and its natural ability for streaming \cite{He19-SET, Zhang20-TTA}.

Following this trend, many open-source speech recognition toolkits now support this model and provide various architectures and features for training and decoding, with an emphasis on different aspects. Tencent’s Pika for example improves the training procedure by introducing Minimum Bayes Risk training and explores composition with architectures such as Causal convolution and TDNN \cite{Weng20-MBR}. RETURNN on the other hand focuses on variant topologies for RNN-T such as RNA \cite{Sak17-RNA} and makes corresponding training tools available \cite{Zeyer18-RGF}. NVIDIA’s NeMo \cite{Kuchaiev19-NTB} includes state-of-art architectures alongside QuartzNet \cite{Kriman20-QND} and also provides streaming examples in order to enable production-level ASR. In the same vein, TensorflowASR \cite{Nguyen20-TFA} capitalizes on the recent advances in the field and offers new state-of-art architectures such as Transformer \cite{Dong18-STN} and Conformer \cite{Gulati20-CCA} and also extensively works on streaming techniques. Finally, recently proposed SpeechBrain \cite{Ravanelli21-SB} acts as an all-in-one toolkit for speech processing and proposes generic tools to train and decode transducer models but also provides many networks, in particular Quaternion \cite{Parcollet18-QCN}.

All of them provide strong tools to reach state-of-art performance or improve decoding speed. However, some important aspects of transducer models are not addressed, such as the unrestricted label expansion during training and decoding \cite{Graves12-STW, Mahadeokar2021-ARS, Tripathi19-MRN}, where an unknown number of labels can be emitted at each timestep. This specific behavior removes vectorization capability and can greatly impact decoding speed and performance \cite{Jain19-RLC, Saon20-ALS}. Additionally, none of them provide tools to improve the generalization ability of the transducer models via auxiliary learning \cite{Jadeberg16-RLU, Toshniwal17-MLL, Liu20-IRN}, which could also help restricting the search space during training without applying strong constraints such as monotonic alignements \cite{Tripathi19-MRN}.\\

In this work, we present an extension of ESPnet \cite{Watanabe18-ESP}, developed to accelerate the research related to this particular model. The extension not only supports composition with architectures such as TDNN, Transformer or Conformer, but also many training and decoding tools. Table~\ref{tab:toolkit_comparison} summarizes the features in the ESPnet toolkit against the mentioned open-source toolkits.  To address the described issues, we focus on introducing and investigating two newly proposed features in the toolkit which are missing from other toolkits: 1) multi-task learning with several auxiliary tasks to improve performance and also reduce the number of expansions at each timestep during training and 
2) various beam search strategies, including our own proposition called N-step Constrained Beam Search, to control the expansion behavior during decoding and also enable different optimizations such as vectorization.
In addition, we will release all configurations, recipes and pre-trained models to the community through ESPnet so that everyone can reproduce our experiments and accelerate the research related to transducer models.

\begin{table*}[tbp]
    \scriptsize
    \centering
    \caption{Comparison with other open-source toolkits supporting transducer models, where "$\star$" represents the ongoing status and "$(\textrm{x})$" denotes the number of supported techniques for a category.}
    \begin{threeparttable}
    \begin{tabular}{l | c c c c c c}
         & NeMo & pika & RETURNN & Speechbrain & TensorflowASR & ESPnet\\
         RNN & \checkmark & \checkmark & \checkmark & \checkmark & \checkmark & \checkmark\\
         Transformer & \checkmark & \checkmark & \checkmark & \checkmark & \checkmark & \checkmark\\
         Conformer & \checkmark & & & \checkmark & \checkmark &\checkmark\\
         Quaternion & & & & \checkmark & &\\
         QuartzNet & \checkmark & & & & &\\
         TDNN encoder & & \checkmark & & \checkmark & & \checkmark\\
         Causal-conv1d decoder & & \checkmark & & & & \checkmark\\
         \hline
         Support CPU training? & \checkmark & \checkmark & \checkmark & & \checkmark & \checkmark\\
         Support discriminative training criterions? & \quad\checkmark (1) & \quad\checkmark (2) & \quad \checkmark (1) & \quad\checkmark (1) & & \quad $\star$ (1)\\
         Support free-form architecture? & & & \checkmark & \checkmark & &\checkmark\\
         Support layer-by-layer parametrization? & \checkmark & & \checkmark & & & \checkmark\\
         Support auxiliary tasks?  & & & & & & \checkmark\\
         Support various beam search decoding strategies? & \quad \,  \, \checkmark (3*) & & & & & \quad\checkmark (4)\\ 
         Support streaming (online decoding)? & \checkmark & & $\star$ & $\star$ & \checkmark & $\star$
    \end{tabular}
    \begin{tablenotes}
    \item[*] Borrowed from ESPnet. See mention in \textbf{BeamRNNTInfer} class\\
    \end{tablenotes}
    \end{threeparttable}
    \label{tab:toolkit_comparison}
    \vspace{-0.6cm}
\end{table*}

\section{Transducer}

The transducer architecture proposed by Graves \cite{Graves12-STW} consists of an encoder, a decoder and a joint network. The encoder, analogous to an acoustic model, encodes a sequence of acoustic features vectors $x$ of length $T$ into a high representation $h^{\textrm{enc}}_{t}$, where $t \leq T$:
\begin{equation}
    h^{\textrm{enc}}_{t} = \textrm{Encoder}(x_{1:T})\label{eq:encoder}
\end{equation}

The decoder, acting as a language model, produces a high representation $h^{\textrm{dec}}_u$ of length $U$ given its previous emitted label sequence $y_{1:u - 1}$:
\begin{equation}
    h^{\textrm{dec}}_{u} = \textrm{Decoder}(y_{1:u - 1})\label{eq:decoder}
\end{equation}

The joint network combines each representations $h^{\textrm{enc}}_{t}$ and $h^{\textrm{dec}}_{u}$ to compute output logits $h^{\textrm{joint}}_{t, u}$ via a network composed of feed-forward layers and a non-linear function:
\begin{equation}
    h^{\textrm{joint}}_{t, u} = \textrm{Joint}(h^{\textrm{enc}}_{t}, h^{\textrm{dec}}_{u})
\end{equation}

Finally, by applying a Softmax function to the output logits, we can produce the distribution of current target probabilities:
\begin{equation}
    P(y_{t, u}|x_{1:T}, y_{1:u - 1}) = \textrm{Softmax}(h^{\textrm{joint}}_{t, u})
\end{equation}

Given $\mathcal{A}$, the set of all possible alignments $a$ between input $x_{1:T}$ and output $y_{1:U}$ with blank labels ($\varnothing$) included, the loss function of the model can be computed as the following negative log likelihood:
\begin{equation}
    \mathcal{L}_{\textrm{trans}} = -\text{log} \sum_{a \in \mathcal{A}} P(a|x_{1:T})\label{eq:trans_loss}
\end{equation}

$\mathcal{L}_{\textrm{trans}}$ can be minimized using the forward-backward algorithm proposed in \cite{Graves12-STW}.

\vspace{-0.25cm}
\section{Framework}

The transducer architecture in ESPnet follows the same encoder-decoder architecture described in \cite{Watanabe18-ESP} used for joint CTC-Attention. Here, each part of the architecture, excluding the joint network, is separated into two sub-parts. Thus, Eq.~\ref{eq:encoder} is replaced by Eq.~\ref{eq:full_encoder} for the encoder and Eq.~\ref{eq:full_decoder} is used instead of Eq.~\ref{eq:decoder} for the decoder.
\begin{equation}
    \begin{aligned}\label{eq:full_encoder}
        &h^{\textrm{pre}}_{t'} = \textrm{EncPre}(x_{1:T})\\
        &h^{\textrm{enc}}_{t} = \textrm{EncBody}(h^{\textrm{pre}}_{t'})
    \end{aligned}
\end{equation}
\vspace{-0.12cm}
\begin{equation}
    \begin{aligned}\label{eq:full_decoder}
        h^{\textrm{dec}}_{y_{u - 1}} &= \textrm{DecPre}(y_{u - 1})\\
        h^{\textrm{dec}}_{u} &= \textrm{DecBody}(y_{u - 1})
    \end{aligned}
\end{equation}
$\textrm{EncPre}(\cdot)$ can be either a 2-layer CNN \cite{Dong18-STN} or a VGG-like max pooling \cite{Lu16-TRN}. $\textrm{DecPre}(\cdot)$ can be either an embedding layer or a linear layer. Following recent advances in ESPnet, various architectures are made available for $\textrm{EncBody}(\cdot)$ and $\textrm{DecBody}(\cdot)$ such as: RNN and variants \cite{Watanabe18-ESP}, Transformer \cite{Karita19-CST}, Transformer with lightweight and dynamic convolution \cite{Fujita20-ABA}, or Conformer \cite{Guo20-RDE}.

Additionally, we introduce what we call \textit{free-form architecture definition}. Here, every components and parameters of the transducer model architecture can be defined and tuned individually. Compared to other models in ESPnet, this also allows us to freely combine previously presented neural networks together or with additional ones (e.g.: Linear, TDNN and Causal-Conv1d \cite{Weng20-MBR}) to form a new architecture for $\textrm{EncBody}(\cdot)$ and $\textrm{DecBody}(\cdot)$.

\section{Augmented training}

Alongside the standard training procedure, we propose an augmented procedure based on various auxiliary tasks. This section describes the ones made available in the ESPnet toolkit.

\subsection{Multi-task learning}

\begin{figure}[tbp]
    \caption{Transducer architecture with auxiliary tasks. In black, the original structure. In red, the new losses, connexions and intermediate networks.}
    \centering
    \includegraphics[width=0.35\textwidth]{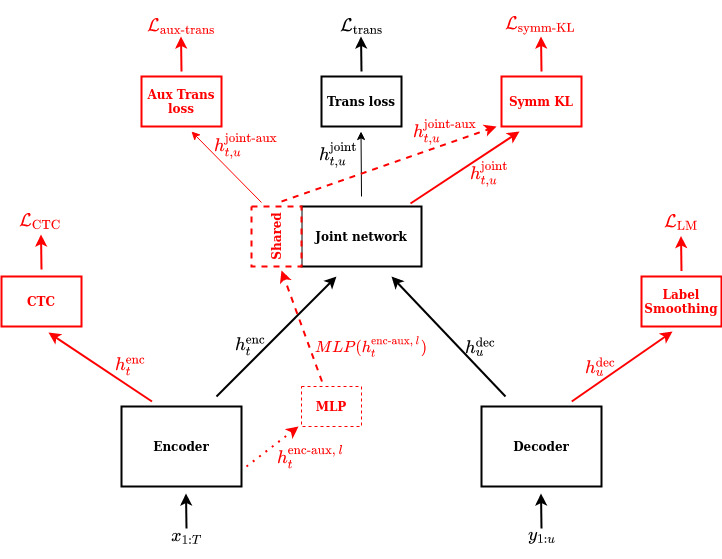}
    \label{fig:augmented_transducer}
\end{figure}

The new transducer structure is augmented by four classifier layers used to train auxiliary tasks alongside the standard transducer criterion. 
The proposed architecture is depicted in Fig. \ref{fig:augmented_transducer}, where a single encoder layer is used to compute $\mathcal{L}_{\textrm{aux-trans}}$ and $\mathcal{L}_{\textrm{symm-KL}}$ (See explanations in Sec. \ref{sec:aux_rnnt} and \ref{sec:aux_symm_kl}).
The five losses can be simultaneously trained and jointly optimize the total loss function $\mathcal{L}_{\textrm{tot}}$ defined as:
\begin{align}\label{eq:final_loss}
    \mathcal{L}_{\textrm{tot}} =\, &\lambda_{\textrm{trans}}\,\mathcal{L}_{\textrm{trans}}\, + \, \lambda_{\textrm{CTC}}\,\mathcal{L}_{\textrm{CTC}}\, + \, \lambda_{\textrm{aux-trans}}\,\mathcal{L}_{\textrm{aux-trans}}\\
    &+ \, \lambda_{\textrm{symm-KL}} \mathcal{L}_{\textrm{symm-KL}}\, +\, \lambda_{\textrm{LM}}\,\mathcal{L}_{\textrm{LM}}\nonumber
\end{align}

where $\mathcal{L}_{\textrm{trans}}$ is the main transducer loss, $\mathcal{L}_{\textrm{CTC}}$ the CTC loss, $\mathcal{L}_{\textrm{aux-trans}}$ the auxiliary transducer loss, $\mathcal{L}_{\textrm{symm-KL}}$ the symmetric KL-divergence and $\mathcal{L}_{\textrm{LM}}$ the LM loss. $\lambda_{x}$ defines their respective contribution to the overall loss. Additionally, each loss can be independently selected or omitted depending on the task.

\subsection{CTC loss ($\mathcal{L}_{\textrm{CTC}}$)}

Similarly to monotonic RNN-T \cite{Tripathi19-MRN} limiting the number of emitted labels at each timestep to strictly one, we explore the use of another (soft) regularization for transducer model through the auxiliary CTC \cite{Jeon20-MLJ}. In addition to the previously found successful encoder pre-initialization with the CTC model \cite{Rao17-EAD}, here, we jointly train the transducer loss and the CTC loss in the same manner as joint CTC-Attention \cite{Kim17-JCA} in ESPnet.

\subsection{Auxiliary transducer loss ($\mathcal{L}_{\textrm{aux-trans}}$)}\label{sec:aux_rnnt}

To address encoder underfitting due to major role given by the decoder in a transducer model, we incorporate the auxiliary transducer loss proposed by Liu et al. \cite{Liu20-IRN} to increase the gradients signals. Following their proposed method, one or multiple encoder representations $h^{\textrm{enc-aux}_{l}}_{t}$, from intermediate layers $l \in L$, are passed to an auxiliary RNN-T criterion, where an additional MLP network is used in place of the linear layer in the joint network:

\begin{equation}
    h^{\textrm{joint-aux}_{l}}_{t, u} = \textrm{JointAux}(\textrm{MLP}(h^{\textrm{enc-aux}_{l}}_{t}), h^{\textrm{dec}}_{u})
\end{equation}

Contrary to the original proposition we choose, based on early experiments, to use an independent joint network we train exclusively for the auxiliary task. However, we do not update the shared decoder parameters and joint parameters if the gradients are back propagated from the auxiliary RNN-T loss.

\subsection{Symmetric KL-divergence ($\mathcal{L}_{\textrm{symm-KL}}$)}\label{sec:aux_symm_kl}

We also consider the use an auxiliary symmetric KL-divergence criterion proposed in \cite{Liu20-IRN} to penalize inconsistent gradients during supervision with auxiliary transducer loss.
\begin{align}
\begin{split}
    \hspace{-0.3cm}\mathcal{L}_{\textrm{symm-KL}} &= \frac{1}{T}\sum_{t=1}^{T}\sum_{l=1}^{L}\frac{1}{U}\sum_{u=1}^{U} I(l)[\\
    \frac{1}{2}(&D_{\textrm{KL}}(\textrm{Softmax}(h^{\textrm{joint}}_{t, u}), \textrm{Softmax}(h^{\textrm{joint-aux}_{l}}_{t, u})) +\\
    &D_{\textrm{KL}}(\textrm{Softmax}(h^{\textrm{joint-aux}_{l}}_{t, u}), \textrm{Softmax}(h^{\textrm{joint}}_{t, u}))],
\end{split}
\end{align}
where $L$ defines the set of encoder layers and $I(l)$ is a binary indicator to denote whether a layer is used for the auxiliary task.

\vspace{-0.2cm}
\subsection{LM loss ($\mathcal{L}_{\textrm{LM}}$)}

Finally, we explore the use of an auxiliary criterion in order to improve and to regularize the decoder network playing a major role in predicting $y_{u}$. Because the decoder is analogous to a language model, we define auxiliary criterion as a language model criterion, where the final computed loss is based on cross-entropy loss with label smoothing \cite{Szegedy16-RTI}. This addition was also proposed in \cite{Jeon20-MLJ} and also used to optimize the decoder independently.

\section{Decoding}\label{sec:decoding}

We support four beam search decoding strategies for transducer to allow more flexibility in regards to performance-speed trade-off and the target task. The first one, our default algorithm (Sec. \ref{subsec:default_bs}), expands the label sequence in a unrestricted manner, the second one (Sec. \ref{subsec:alsd}) acts label-synchronously and the last two strategies (Sec. \ref{subsec:tsd} and Sec. \ref{subsec:nsc}) runs along time axis. 
For three of them, we enable hypotheses caching and batching to reduce the computation time. Shallow fusion with a RNN-based or Transformer-based LM and multi-level LM \cite{Hori17-MLL} is also supported.

\subsection{Default beam search}\label{subsec:default_bs}

The default decoding strategy for transducer in ESPnet is based on the beam search algorithm proposed by Graves in \cite{Graves12-STW}. The procedure runs alongs both axis $\{1, ..., T\}$ and $\{1, ..., U\}$ and can be expanded in an unrestricted manner.
The algorithm is performed using two sets of hypotheses A and B, respectively the set of hypotheses for times $t$ and $t + 1$, where B contains $\varnothing$ hypothesis at $t = 0$. At each timestep $t$, hypotheses from $B$ are first moved to $A$. The best hypothesis from $A$ is extracted and expanded with either a blank transition or non-blank transition. Hypotheses ending with a blank transition are stocked in $B$ whereas the others are moved to $A$, where each hypothesis' score is updated with the corresponding blank or non-blank transition score. Through a while loop, the procedure is repeated until $B$ contains at least $N_{\textrm{bs}}$ hypotheses more probable than the most probable in $A$. When the condition is met, $B$ is pruned to $N_{\textrm{bs}}$ hypotheses and passed to the next timestep. At the end of the procedure when $t = T$, the N-best hypotheses in $B$ are returned sorted by descending score.

Based on initial experiments, the prefix search part was removed. We found out that it does not necessary insure that the search space won't be redundant without additional duplication check and, in the end, most probable hypotheses are still retained. Thus, removing that part allows us to reduce the computational cost without trading off much of the search accuracy.

\vspace{-0.2cm}
\subsection{Alignment-length synchronous decoding (ALSD)}\label{subsec:alsd}
\vspace{-0.3cm}
\begin{algorithm}
\DontPrintSemicolon
\caption{\footnotesize Alignment-length synchronous decoding \cite{Saon20-ALS}}
\label{algo:alsd}
\textbf{Input: } $h^{\textrm{enc}}_{1:T}$, $U_{\textrm{max}}$, $N_{\textrm{bs}}$ and $N_{\textrm{best}}$\\
 $B = \{\varnothing, 1, \textrm{state}^{\textrm{dec}}_{0}\}$; $F$ = \{\}\;
 \For{i =  1 ... T + $U_{\textrm{\upshape max}}$}{
    $A$ = \{\}\;
    \For{$(y, \delta_{i-1}(y), \textrm{\upshape state}^{\textrm{\upshape dec}}_{u-1}) \in$ B}{
        $u = |y|$\;
        t = i - u + 1\;
        \If{t $>$ T}{
            continue\;
        }
        $h^{\textrm{dec}}_{u}, \textrm{state}^{\textrm{dec}}_{u} = \textrm{Decoder}(y, \textrm{state}^{\textrm{dec}}_{u-1})$\;
        $p^{\textrm{pr}} = \textrm{Softmax}(\textrm{Joint}(h^{\textrm{enc}}_{t}, h^{\textrm{dec}}_{u}))$\;
        $\delta_{i}(y) = \delta_{i-1}(y) * p^{\textrm{pr}}(\varnothing)$\;
        $A = A \cup \{(y, \delta_{i}(y), \textrm{state}^{\textrm{dec}}_{u-1})\}$\;
        \If{t == T}{
            $F = F \cup \{(y, \delta_{i}(y))\}$\;
        }
        \For{k $\in$ Y}{
            $\delta_{i}(y + k) = \delta_{i-1}(y) * p^{\textrm{pr}}(k)$\;
            $A = A \cup \{(y + k, \delta_{i}(y + k), \textrm{state}^{\textrm{dec}}_{u})\}$\;
        }
    }
    $B$ = PruneAndRecombineHyps($A$)[:$N_{\textrm{bs}}$]\;
 }
 \textbf{Return: } SortedByScore($F$)[:$N_{\textrm{best}}$]\;
\end{algorithm}

Alignment-length synchronous decoding is the procedure proposed by Saon et al. \cite{Saon20-ALS} which runs along axis $\{1, ..., U\}$ and uses $U_\textrm{max}$ parameter, an estimate of the maximum output sequence length, where $U_\textrm{max} < T$.
The procedure keeps track of 2 set of hypotheses $A$ and $B$ at alignment step $i$ and $i - 1$. At step $i$, for each hypothesis of $B$, the number of frames covered by the output sequence $y$ is computed by subtracting its length from $i + 1$, then the hypothesis is added to $A$ with its score ($\delta_{x}$) updated by adding the blank transition score. If the last frame is reached, the hypothesis is also put into the set of final hypotheses $F$. After that, each $y$ of $A$ are expanded with every output label minus blank, and each new hypothesis is added to $A$ with its corresponding score and decoder network state updated. Finally, a pruning is applied for the set $A$ and duplicate hypotheses are merged altogether with their respective score added. The set of unique hypotheses, reduced to the beam size ($N_{bs}$), become the set $B$ for the next alignment step. At the end of the procedure, the N-best ($N_{\textrm{best}}$) hypotheses in $F$ are returned sorted by descending score. The complete procedure is given in Algo. \ref{algo:alsd}.

\vspace{-0.1cm}
\subsection{Time-synchronous decoding (TSD)}\label{subsec:tsd}
\begin{algorithm}
\DontPrintSemicolon
\caption{Time synchronous decoding \cite{Saon20-ALS}}
\label{algo:tsd}
\textbf{Input: } $h^{\textrm{enc}}_{1:T}$, max\_sym\_exp, $N_{\textrm{bs}}$ and $N_{\textrm{best}}$\;
 $B = \{\varnothing, 1, \textrm{state}^{\textrm{dec}}_{0}\}$\;
 \For{t =  1 ... T}{
    $A$ = \{\}; $C$ = $B$\;
    \For{v = 1 ... $\textrm{\upshape max\_sym\_exp}$}{
        $D$ = \{\}\;
        \For{$(y, \delta_{t-1, u-1}(y), \textrm{\upshape state}^{\textrm{\upshape dec}}_{u-1}) \in \textrm{C}$}{
             $h^{\textrm{dec}}_{u}, \textrm{state}^{\textrm{dec}}_{u} = \textrm{Decoder}(y, \textrm{state}^{\textrm{dec}}_{u-1})$\;
             $p^{\textrm{pr}} = \textrm{Softmax}(\textrm{Joint}(h^{\textrm{enc}}_{t}, h^{\textrm{dec}}_{u}))$\;
             
             \eIf{$y \notin A$}{
                $\delta_{t, u - 1}(y) = \delta_{t-1, u-1}(y) * p^{\textrm{pr}}(\varnothing)$\;
                $A = A \cup \{(y, \delta_{t, u-1}(y), \textrm{state}^{\textrm{dec}}_{u-1})\}$\;
             }{
                $\delta_{t - 1, u}(y) = \delta_{t-1, u-1}(y) * p^{\textrm{pr}}(k)$\;
             }
             
             \If{v $< \textrm{\upshape max\_sym\_exp}$}{
                \For{k $\in$ Y}{
                    $\delta_{t - 1, u}(y + k) = \delta_{t-1, u-1}(y) * p^{\textrm{pr}}(k)$\;
                    \footnotesize $D = D \cup \{(y + k, \delta_{t-1, u}(y + k), \textrm{state}^{\textrm{dec}}_{u})\}$\;\small
                }
             }
        }
        $C$ = PruneHyps($D$, $N_{\textrm{bs}}$)\;
    }
    $B$ = $A$\;
 }
 \textbf{Return: } SortedByScore(B)[:$N_{\textrm{best}}$]\;
\end{algorithm}

Time-synchronous decoding is a procedure also proposed by Saon et al. \cite{Saon20-ALS}. It runs along axis $\{1, ..., T\}$ and uses a parameter $\textrm{max\_sym\_exp}$ to control the number of hypotheses expansion at each timestep. Algorithm \ref{algo:tsd} shows the complete procedure and can be described as follow.
Here, the procedure keep track of 4 set of hypotheses, where $A$ and $B$ store hypotheses for times $t$ and $t - 1$ and $C$ and $D$ store hypotheses for expansion steps $v - 1$ and $v$. At each timestep, in case of a blank transition, hypotheses from $C$ are added to $A$ with blank score added to their score if $y \not\in A$, otherwise scores (($\delta_{x}$) are summed. For a non-blank transition, hypotheses from $C$ are expanded with every output label minus blank and added to set $D$ with their score updated. After that, hypotheses from $D$ are pruned to define a new set $C$ limited to $N_{\textrm{bs}}$ hypotheses. The procedure is then repeated $\textrm{max\_sym\_exp}$ times. When $v$ reaches $\textrm{max\_sym\_exp}$, hypotheses from $A$ are stored in $B$ and the procedure is repeated for each timestep. At the end of the procedure, the N-best ($N_{\textrm{best}}$) hypotheses in $B$ are returned sorted by a descending score.

\subsection{N-step constrained beam search (NSC)}\label{subsec:nsc}

We also include an improved version of the One-Step Constrained (OSC) beam search proposed by Kim et al. \cite{Kim20-ART}, which originally constrains the default beam search to a single label emission plus blank label at each timestep.
Although the authors demonstrate the efficiency of the algorithm for different speech recognition tasks and investigate the number of emitted labels at each timestep during the expansion search for the presented tasks, we found the initial constraint too strong, resulting in two weakness:

\begin{algorithm}[!ht]
\DontPrintSemicolon
\caption{N-step constrained beam search.}
\label{algo:nsc}
\textbf{Input: } $h^{\textrm{enc}}$, $N_\textrm{step}$, $\textrm{auto-}N_{\textrm{step}}$, $\alpha$, $N_{\textrm{bs}}$ and $N_{\textrm{best}}$\\
 $h^{\textrm{dec}}_{\textrm{batch}}, \textrm{state}^{\textrm{dec}}_{\textrm{batch}} = \footnotesize \textrm{BatchDecoder}(\varnothing, \textrm{Duplicate}(\textrm{state}^{\textrm{dec}}_{0}, N_{\textrm{bs}}))$\;\small
 $B$ = $\{\varnothing, 1, h^{\textrm{dec}}_{\textrm{batch}_{0}}, \textrm{state}^{\textrm{dec}}_{\textrm{batch}_{0}}\}$\;
 \For{t =  1 ... T}{
    $A$ = SortedByLength($B$); $B$ = \{\}\;
    \For{$(y_{i}, \delta(y_{i})) \in$ A}{
        $\delta(y_{i})\, \text{+=} \sum_{\hat{y}} \delta(\hat{y}) * \delta(y_{i} | \hat{y}, t)$\;
        \, \textbf{where} $\hat{y} \in pref(y_{i}) \cap A \text{ and } |y_{i}| - |\hat{y}| < \alpha$\;
    }
    $S$ = \{\}; $V$ = \{\}\;
    \begingroup
    \color{red}
    \For{n = 1 ... $N_\textrm{\upshape step}$}{
        \begingroup
        \color{black}
        $h^{\textrm{dec}}_{A} = \{h^{\textrm{dec}}_{y_{1}}, ..., h^{\textrm{dec}}_{y_{N_{\textrm{bs}}}}\} \in A$\;
        $p^{\textrm{pr}} = \textrm{Softmax}(\textrm{Joint}(h^{\textrm{enc}}_{\textrm{t}}, h^{\textrm{dec}}_{A}))$\;

        \For{$(y_{i}, \delta(y_{i}), h^{\textrm{\upshape dec}}_{y_{i}}, \textrm{\upshape state}^{\textrm{\upshape dec}}_{y_{i}}) \in A$}{
            $\delta(y_{i}) = \delta(y_{i}) * p^{\textrm{pr}}(\varnothing)$\;
            $S = S \cup \{(y_{i}, \delta(y_{i}), h^{\textrm{dec}}_{y_{i}}, \textrm{state}^{\textrm{dec}}_{y_{i}})\}$\;

            \For{k $\in$ Y}{
                $\delta(y_{i} + k) = \delta(y_{i}) * p^{\textrm{pr}}(k)$\;
                $V = V \cup \{(y_{i} + k, \delta(y_{i} + k), h^{\textrm{dec}}_{y_{i}}, \textrm{state}^{\textrm{dec}}_{y_{i}})\}$\;
            }
        }
        $V$ = SubstractSet(SortedByScore($V$), $B$)[:$N_{\textrm{bs}}$]\;
        $y_{\textrm{batch}} = \{y_{1}, ..., y_{N_{\textrm{bs}}}\} \in V$\;
        $\textrm{state}^{\textrm{dec}}_{\textrm{batch}} = \{\textrm{state}^{\textrm{dec}}_{y_{1}}, ..., \textrm{state}^{\textrm{dec}}_{y_{N_{\textrm{bs}}}}\} \in V$\;
        $h^{\textrm{dec}}_{\textrm{batch}}, \textrm{state}^{\textrm{dec}}_{\textrm{batch}} = \textrm{BatchDecoder}(y_{\textrm{batch}}, \textrm{state}^{\textrm{dec}}_{\textrm{batch}})$\;
        \endgroup
        \eIf{$n < N_{\textrm{\upshape step}} - 1$}{
            \For{$(y_{i}, h^{\textrm{\upshape dec}}_{y_{i}}, \textrm{\upshape state}^{\textrm{\upshape dec}}_{y_{i}}) \in V$}{
                $h^{\textrm{dec}}_{y_{i}} = h^{\textrm{dec}}_{\textrm{batch}_{i}}$\;
                $\textrm{state}^{\textrm{dec}}_{y_{i}} = \textrm{state}^{\textrm{dec}}_{\textrm{batch}_{i}}$\;
            }
            $A$ = $V$\;
        }{
            \begingroup
            \color{black}
            $p^{\textrm{pr}} = \textrm{Softmax}(\textrm{Joint}(h^{\textrm{enc}}_{\textrm{t}}, h^{\textrm{dec}}_{\textrm{batch}}))$\;
            \For{$(y_{i}, \delta(y_{i}), h^{\textrm{\upshape dec}}_{y_{i}}, \textrm{\upshape state}^{\textrm{\upshape dec}}_{y_{i}}) \in V$}{
                \begingroup
                \color{red}
                \If{$N_{\textrm{step}} = 1$ and $\textrm{auto-}N_{\textrm{step}} = 1$}{
                    $\delta(y_{i}) = \delta(y_{i}) * p^{\textrm{pr}}(\varnothing)$\;
                }
                \endgroup
                $h^{\textrm{dec}}_{y_{i}} = h^{\textrm{dec}}_{\textrm{batch}_{i}}$\;
                $\textrm{state}^{\textrm{dec}}_{y_{i}} = \textrm{state}^{\textrm{dec}}_{\textrm{batch}_{i}}$\;
            }
            \endgroup
        }
    }
    \endgroup
    $B$ = SortedByScore($S$ + $V$)[:$N_{\textrm{bs}}$]\;
 }
 \textbf{return: } SortedByScore($B$)[:$N_{\textrm{best}}$]\;
\end{algorithm}

\begin{enumerate}
\item For low-resources tasks, the number of needed label emissions at each timestep should be higher than 1 as shown Table \ref{tab:trans_expansion}. Here, the investigation on expansion search was conducted on two smaller corpora, VIVOS (15 hours) and Voxforge (20 hours), and a significant number of more than one expansions was observed in comparison to the initial investigation (+4.24\% for Voxforge and +7.28\% for VIVOS).
\item If no equivalent constraint is applied during training (e.g.: \cite{Tripathi19-MRN}), adding a blank transition score after each expansion may result in the deletion of reasonable hypotheses during the pruning process and sorting phase.
\end{enumerate}

\begin{table}
    \centering
    \caption{Number of expansions (\%) during expansion search.}
    \begin{tabular}{c|c|c|c}
         & VIVOS &  Voxforge & TIMIT \cite{Kim20-ART}\\
         \hline
         Num. exp & \% exp. & \% exp. & \% exp.\\
         \hline
         1 & 89.62 & 93.18 & 97.73\\
         2 & 9.52 & 6.48 & 2.24\\
         3 & 0.86 & 0.34 & 0.03\\
    \end{tabular}
    \label{tab:trans_expansion}
\end{table}

In order to address these issues, we propose a novel \textit{N-Step Constrained beam search} (NSC) algorithm. The algorithm is similar to the TSD (Sec. \ref{subsec:tsd}) and extends the original OSC algorithm to $N$ expansion steps (plus blank) through an additional loop controlled by a parameter $N_\textrm{step}$. To overcome the second issue, we add a new condition for the final expansion step in NSC. If $N_\textrm{step}=1$ and $\textrm{auto-}N_\textrm{step}>1$, then we allow incomplete hypotheses to be passed to the next time step without adding the blank score transition. The parameter $\textrm{auto-}N_\textrm{step}$ is obtained by a counting method, built upon default beam search (See Sec. \ref{subsec:default_bs}), which computes the expected number of needed expansions. The complete procedure is given by Algorithm \ref{algo:nsc}, where lines highlighted in red refer to our addition to the original OSC algorithm, and the parameters $N_{\textrm{bs}}$ and $N_{\textrm{best}}$ define respectively the beam size and the $N$ best hypotheses.

With the exception of default beam search, all strategies propose to resolve the unrestricted expansion in transducer model by controlling the search space either time-synchronously (TSD, OSC NSC) or label-synchronously (ALSD). While the latter was found promising in regards to its decoding speed, it cannot be used for streaming \cite{Saon20-ALS}. For the strategies running along time axis, OSC and TSD shown some drawbacks compared to our proposed algorithm, which can be summarized as follow: 1) OSC rely on a too strong constraint resulting in performance degradation for some tasks such as low resources and 2) TSD, while adding more control to alleviate previous issue, fell short against our own due to the blank transition score addition on some edge cases (see Sec. \ref{sec:invest_decoding}.)
\vspace{-0.15cm}
\section{Experiments}

Our experiments are formulated as follow. First, we investigate the proposed training augmentation through several experiments. Second, we compare the decoding strategies in regards to both error rate and real-time factor. Finally, we present a comparison against other models in ESPnet and state-of-art systems on different datasets.
For the first two sections, we use the Voxforge italian dataset \cite{Voxforge19} ($\sim$ 20 hours), which has the advantages of being free and experiments are easily reproducible without a consequent need in terms of resources.
For the last section, we use LibriSpeech \cite{Panayotov15-ACB} and AISHELL-1 \cite{Bu17-AOS}. For the different configurations and the pre-trained models used in our experiments , we refer the reader to the ESPnet toolkit where all needed resources will be made available.

\vspace{-0.15cm}
\subsection{Architecture and training augmentation}

To assess the effectiveness of the auxiliary tasks, we conduct several experiments with a RNN-T model on the Voxforge dataset. In order to perform our primary experiments, we first need to rank our auxiliary tasks, based on the observed average CER/WER gain, and select an optimal weight for each task. Thus, we performed an empirical study where we test every possible weights  for each task (i.e.: $0<\lambda_x<1.0$ in Eq.~\ref{eq:final_loss}) under two training conditions: 1) with a single auxiliary task and 2) with every combination of auxiliary tasks. From here, our final ranking is as follows: 1) $\mathcal{L}_{\textrm{CTC}}$ with $\lambda_{\textrm{CTC}}=0.5$, 2) $\mathcal{L}_{\textrm{LM}}$ with $\lambda_{\textrm{LM}}=\{0.4,0.5\}$, 3) $\mathcal{L}_{\textrm{aux-trans}}$ with $\lambda_{\textrm{aux-trans}}=0.3$ coupled to $\mathcal{L}_{\textrm{symm-KL}}$ with $\lambda_{\textrm{symm-KL}}={0.2}$, and finally $\mathcal{L}_{\textrm{aux-trans}}$ alone with the same weight. While we observed an average gain in terms of CER and WER with $\mathcal{L}_{\textrm{CTC}}$ and  $\mathcal{L}_{\textrm{LM}}$, $\mathcal{L}_{\textrm{aux-trans}}$, paired or not with  $\mathcal{L}_{\textrm{symm-KL}}$, was found beneficial only with specific weights and encoder layer(s) connected.

\begin{table}[tbp]
  \caption{Ablation results on the dev and test sets of Voxforge. CER and WER (in \%) are reported with the default beam search.}
  \label{tab:ablation}
  \centering
    \begin{tabular}{ l c c c c}
     \toprule
     \textbf{Model} & \textbf{CER} & \textbf{WER} &\textbf{CER} & \textbf{WER} \\
     \midrule
     \textbf{Attention} \\
     \quad RNN \cite{Karita19-CST} & 12.9 & n.a & 12.6 & n.a\\
     \quad Conformer \cite{Guo20-RDE} & \textbf{8.7} & n.a & \textbf{8.2} & n.a\\
     \hline
     \textbf{Transducer (ours)} \\
     \quad Augmented RNN-T & \textbf{12.2} & \textbf{41.7} & \textbf{11.5} & \textbf{40.4}\\
     \quad \quad – $\mathcal{L}_{\textrm{symm-KL}}$ & \textbf{12.2} & 42.0 $\uparrow$ & \textbf{11.5} & 40.5 $\uparrow$\\
     \quad \quad \quad – $\mathcal{L}_{\textrm{aux-trans}}$ & 12.1 $\downarrow$ & 42.3 $\uparrow$ & 11.4 $\downarrow$ & 41.0 $\uparrow$\\
     \quad \quad \quad \quad – $\mathcal{L}_{\textrm{LM}}$ & 12.5 $\uparrow$ & 43.9 $\uparrow$ & 11.6 $\uparrow$ & 42.1 $\uparrow$\\
     \quad \quad \quad \quad \quad – $\mathcal{L}_{\textrm{CTC}}$ & 13.1 $\uparrow$ & 44.4 $\uparrow$ & 12.2 $\uparrow$ & 42.4 $\uparrow$\\
     \bottomrule
    \end{tabular}
\end{table}

For our first primary analysis, we perform an ablation study on the RNN-T system using all auxiliary task to understand how each task contributes to the overall task. The results of our study are reported in Table \ref{tab:ablation}, where task are removed by descending order based on the initial ranking. Overall, most auxiliary tasks contribute to the main task with the exception of $\mathcal{L}_{\textrm{\textrm{aux-trans}}}$ adding errors at character level to improve WER and $\mathcal{L}_{\textrm{\textrm{symm-KL}}}$ slightly improving the WER only. Most improvement come from the addition of $\mathcal{L}_{\textrm{\textrm{CTC}}}$ and $\mathcal{L}_{\textrm{\textrm{LM}}}$, with a CER/WER decrease on the dev and test set of respectively -0.6\%/-0.5\% and -0.6\%/-0.3\% for the first, and -0.5\%/-1.6\% and -0.2\%/-1.1\% for the latter. While the auxiliary CTC addition mostly decreases the number of errors at character level by enforcing monotonicity, adding LM criterion further improves decoder prediction resulting in a significative gain in terms of the WER.

\begin{table}[tbp]
  \caption{Comparaison of auxiliary tasks against transfer learning (p.t.) and decoding with ext. LM. CER and WER (in \%) are reported on Voxforge dev and test set with the default beam search.}
  \label{tab:focus_ctc_lm}
  \centering
    \begin{tabular}{ l c c c c}
     \toprule
     \textbf{Model} & \textbf{CER} & \textbf{WER} &\textbf{CER} & \textbf{WER} \\
     \midrule
     RNN-T & 13.1 & 44.4 & 12.2 & 42.4\\
     RNN-T + $\mathcal{L}_{\textrm{\textrm{CTC}}}$ & 12.6 & 43.9 & 11.6 & 42.1\\
     RNN-T + $\mathcal{L}_{\textrm{\textrm{LM}}}$ & 13.0 & 42.9 & 12.1 & 41.6\\
     RNN-T + $\mathcal{L}_{\textrm{\textrm{CTC}}}$ + $\mathcal{L}_{\textrm{\textrm{LM}}}$ & 12.2 & 42.3 & 11.4 & 41.0\\
     \midrule\midrule
     RNN-T w/ p.t. CTC & 12.3 & 44.5 & 11.2 & 42.3\\
     \midrule
     RNN-T + ext. LM & 12.8 & 42.3 & 12.0 & 40.9\\
     RNN-T w/ p.t. LM & 12.8 & 43.0 & 12.3 & 42.0\\
     \midrule
     RNN-T + $\mathcal{L}_{\textrm{\textrm{CTC}}}$ + ext. LM & 12.4 & 42.2 & 11.5 & 40.6\\
     RNN-T w/ p.t. LM + $\mathcal{L}_{\textrm{\textrm{CTC}}}$ & 12.7 & 44.2 & 12.0 & 43.3\\
     RNN-T w/ p.t. CTC + ext. LM & 12.1 & 42.4 & 11.0 & 40.9\\
     RNN-T w/ p.t. CTC + $\mathcal{L}_{\textrm{\textrm{LM}}}$ & \textbf{11.8} & \textbf{42.1} & \textbf{10.9} & \textbf{40.5}\\
     \bottomrule
    \end{tabular}
\end{table}

Next, we compare our two best performing auxiliary tasks, $\mathcal{L}_{\textrm{CTC}}$ and  $\mathcal{L}_{\textrm{LM}}$, against other techniques incorporating CTC and LM tasks: 1) transfer learning with a pre-trained CTC model for the encoder part \cite{Rao17-EAD} and a pre-trained language model \cite{Hu20-EPA} for the decoder part, and 2) decoding with an external language model. The table \ref{tab:focus_ctc_lm} summarizes our experiments, where "ext. LM" refers to the external LM and "p.t. $X$" to the transfer of pre-trained weights from a model $X$.
Relying on a pre-trained CTC model for encoder initialization results in a lower CER and an higher WER compared to training a transducer model with $\mathcal{L}_{\textrm{CTC}}$. For the LM task, training a vanilla RNN-T and using an external LM for decoding brings the most improvement (avg. -0.15\% CER and -0.9\% WER) at the cost of a significant increase in terms of decoding time (almost doubled). Using transfer learning or $\mathcal{L}_{\textrm{LM}}$ to regularize the decoder part during training brings almost the same performance, with a slightly better performance in terms of CER (-0.2\%) for the first and a WER improvement (-0.25\%) for the latter. From our observation, transfer learning for either part of the model makes the model starts with more emphase on the conditional independence between predictions in comparison to training the model with an auxiliary task.

From here, we extend our investigation with a comparison between different pairs of techniques. Most pairs seems to work well together and further decrease both CER and WER compared to training with a single technique. Notably, we found out that pre-initializing the encoder with a CTC model and adding an auxiliary LM task during training to focus on the decoder part decrease significantly the number of errors at character level compared to training with $\mathcal{L}_{\textrm{CTC}}$ and $\mathcal{L}_{\textrm{LM}}$. That setup also outperforms our model trained with all auxiliary tasks in terms of CER (avg. -0.4\%) but brings more errors at a word level (+0.25\% WER). Additionally, we observed a serious performance degradation for both CER and WER when initializing the decoder with a pre-trained LM and training the model with $\mathcal{L}_{\textrm{CTC}}$. Finally, using an external LM during decoding further improves CER and WER under all conditions.

\subsection{Investigation on decoding strategies}\label{sec:invest_decoding}

\begin{figure}[tbp]
    \centering
    \caption{CER against RTF for the decoding strategies. (1) denotes the baseline model and (2) the model trained with auxiliary tasks.}
    \includegraphics[width=0.45\textwidth,
    height=0.17\textheight]{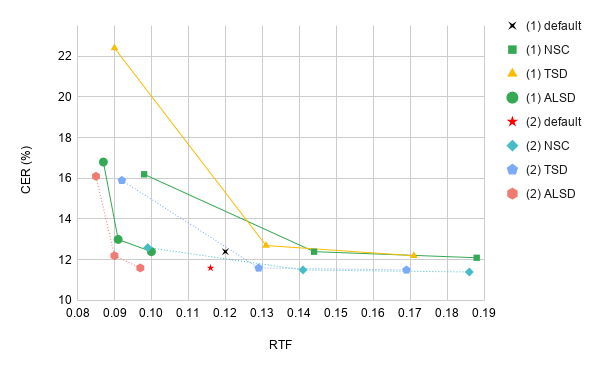}
    \label{fig:beam_search}
    \vspace{-0.2cm}
\end{figure}

Figure \ref{fig:beam_search} compares the decoding strategies, introduced in section \ref{sec:decoding}, with different parameter values in terms of CER vs RTF, averaged on 5 runs. We use the vanilla RNN-T and the RNN-T trained with all auxiliary tasks, and a beam of size 5 under all conditions. Three different values are evaluated for each algorithm: [25, 50, 100] for $U_{\textrm{max}}$ in ALSD, [2, 3, 4] for max\_sym\_exp in TSD and [1, 2, 3] for $N_{\textrm{step}}$ in NSC. All experiments were performed using a CPU Intel i7-6950X limited to one thread (the default setting in ESPnet).\\

With the baseline model, ALSD outperforms all other algorithms in terms of the decoding speed, with a maximum RTF of 0.1, comparable to the best RTF observed for frame-synchronous strategies. For the latter, TSD and NSC, reducing each time by one the value for the parameter controlling number of emitted labels, respectively max\_sym\_exp and $N_{\textrm{step}}$, results in an average 25\% RTF reduction but increase the CER. Because TSD suffers from the same issue as OSC (See \ref{subsec:nsc}), its performance is significantly impacted when $\textrm{max\_sym\_exp}=2$, resulting in an CER increase of 80\% while NSC with $N_\textrm{step}=1$ is increased by 36\%. Reducing $U_\textrm{max}$ for ALSD also significantly impacts the CER (+29\% with $U_\textrm{max}=25$).
Using auxiliary tasks reduces the number of emitted labels by timestep, mainly due to the auxiliary CTC. Compared to the report in Table~\ref{tab:trans_expansion}, we denote the following: 96.32\% of 1-expansion, 3.5\% of 2-expansion and 0.18\% of 3-expansion. The overall RTF is only slightly reduced but it results in a lower CER increase for the frame-synchronous strategies, notably: 37\% and 9.5\% for respectively TSD with $\textrm{max\_sym\_exp}=2$ and NSC with $N_\textrm{step}=1$. For ALSD, the impact on CER and RTF is similar but the strategy with $U_\textrm{max}=25$ is now outperformed in terms of CER by NSC with $N_{\textrm{step}}=1$ (16.1\% vs 12.6\% CER). In regards to the CER/RTF trade-off, NSC with these parameters seems promising in comparison to our default algorithm.

\subsection{Performance comparison}

We evaluate ESPnet's transducer models on AISHELL-1 \cite{Bu17-AOS} and LibriSpeech \cite{Panayotov15-ACB} to compare their performance with other models in the literature. Specifically, RNN-T and Conformer-T models without and with auxiliary tasks are evaluated, where $\mathcal{L}_{\textrm{CTC}}$ is used for AISHELL-1, and $\mathcal{L}_{\textrm{CTC}}$ and $\mathcal{L}_{\textrm{LM}}$ are used for LibriSpeech. The default beam search is used with a beam size of 10 for both datasets.
Table~\ref{tab:aishell} presents the results on AISHELL-1. Our RNN-T and Conformer-T models achieved CERs of 7.2\% and 5.0\% respectively on the test set, which are both significantly better than the performance of RNN-T/Conformer-T models in the literature \cite{Tian19-SAT}\cite{Huang20-IRT}. With the help of the auxiliary task, the RNN-T and Conformer-T models reached CERs of 6.9\% and 4.7\%, respectively. The latter is on par with the state-of-the-art performance achieved by the attention-based Conformer model \cite{Guo20-RDE}.
Table~\ref{tab:librispeech} summarizes the results on LibriSpeech. The use of the auxiliary tasks again helped to reduce the WERs of our RNN-T model on the test sets. An even better performance of 2.9\%/6.8\% was obtained by switching from RNN-T to Conformer-T. Finally by using shallow-fusion and an external LM trained on the default training material for LibriSpeech's LM, the Conformer-T achieved 2.6\%/6.1\%, which is comparable to the performance of other open-source toolkits. 

\begin{table}[tbp]
  \caption{CERs (\%) on the dev/test sets of AISHELL-1.}
  \label{tab:aishell}
  \centering
    \begin{tabular}{ l c c }
     \toprule
     \textbf{Model} & \textbf{Dev} & \textbf{Test} \\
     \midrule
     \textbf{Attention} \\
     \quad RNN \cite{Karita19-CST} & 6.8 & 8.0 \\
     \quad Conformer \cite{Guo20-RDE} & \textbf{4.4} & \textbf{4.7} \\
     \textbf{Transducer} \\
     \quad RNN-T \cite{Tian19-SAT} & 10.1 & 11.8 \\
     \quad Conformer-T \cite{Huang20-IRT} & n/a & 5.4 \\
     \hline
     \textbf{Transducer (ours)} \\
     \quad RNN-T & 6.3 & 7.2 \\ 
     \quad RNN-T + Aux & 6.2 & 6.9 \\  
     \quad Conformer-T & 4.5 & 5.0 \\
     \quad Conformer-T + Aux & \textbf{4.3} & \textbf{4.7} \\
     \bottomrule
    \end{tabular}
\end{table}

\begin{table}[tbp]
  \caption{WERs (\%) on the dev/test sets of LibriSpeech.}
  \label{tab:librispeech}
  \footnotesize  
  \centering
    \begin{tabular}{@{\extracolsep{4pt}} l c c c c @{}}
     \toprule
     \textbf{Model} & \multicolumn{2}{c}{\textbf{Dev}} & \multicolumn{2}{c}{\textbf{Test}} \\
     \cline{2-3} \cline{4-5}
     & \textbf{clean} & \textbf{other} & \textbf{clean} & \textbf{other} \\
     \midrule
     \textbf{Other Open-Source Toolkits} \\
     \quad NeMo (QuartzNet) \cite{Kriman20-QND} & n/a & n/a & 2.69 & 7.25 \\
     \quad SpeechBrain (Transformer) \cite{Ravanelli21-SB} \hspace{-2em} & n/a & n/a & 2.46 & 5.86 \\
     \quad RETURNN (Transducer) \cite{Zeyer21-LTM} & 2.17 & 5.28 & 2.23 & 5.74 \\
     \quad ESPnet (Conformer) \cite{Guo20-RDE} & \textbf{1.9} & \textbf{4.9} & \textbf{2.1} & \textbf{4.9} \\
     \hline
     \textbf{Transducer (ours)} \\
     \quad RNN-T & 4.0 & 11.2 & 4.2 & 11.4 \\ 
     \quad RNN-T + Aux & 3.8 & 10.8 & 4.0 & 10.9 \\ 
     \quad Conformer-T + Aux & 2.7 & 6.7 & 2.9 & 6.8 \\
     \quad \quad + LM shallow fusion & \textbf{2.4} & \textbf{5.9} & \textbf{2.6} & \textbf{6.1} \\
     \bottomrule
    \end{tabular}
    \vspace{0.25cm}
\end{table}


\section{Conclusion}

This paper introduced an extension of the ESPnet speech recognition toolkit dedicated to the transducer models. Through an experimental evaluation of the models and some main proposed features, we demonstrated that our models can achieve state-of-art results on AISHELL-1 dataset and also exhibit promising performance in regards to real-time decoding and others models in ESPnet.
Future work will focus on improving the model trained on Librispeech and adding features currently in development, namely: training with MBR and streaming. Further analysis of our proposed Transducer against non-autoregressive models \cite{Higuchi20-MCN, Higuchi21-IMC} is also considered.

\newpage
\bibliographystyle{IEEEtran}
\newpage
\bibliography{mybib}

\end{document}